# The efficiency and power of the martensite rotor heat engine. I

**I.G. Margvelashvili, M.D. Zviadadze, L.A. Zamtaradze**

I. Javakhishvili Tbilisi State University, 3 Chavchavadze Ave. Tbilisi, 0128, Georgia
E. Andronikashvili Institute of Physics, 6 Tamarashvili Str., Tbilisi, 0177, Georgia

zviadadzemichael@yahoo.com

## Abstract

The physical aspects - mechanics and thermodynamics – of operation of martensite rotor heat engine (MRHE) on the basis of martensite-austenite structural phase transition with the transition temperature in the region of low-potential water temperatures have been studied. The engine converts the thermal energy of low-potential water into the elastic energy of working body (spring, ribbon or wire) made of the material with shape memory effect. At some simplifying assumptions, the analytical expressions are obtained for the thermal efficiency and the power of MRHE of different type. The registration of head hydraulic resistance and heat conductivity of working body material is made and the maximum value of power produced by the engine at the given mechanical and heat conditions is calculated.

The recommendations are given on the optimal choice of engine parameters. On the basis of numerical estimations for nitinol, the possibility of application of MRHE is shown for efficient and ecologically pure production of electric energy both on local (geothermal waters, waste water of industrial enterprises, etc.) and global (warm ocean stream) scales.

# 1. Introduction

The problem of creation of heat engines operating on the basis of structural phase transitions in solids has been of special interest for researchers during several tens of years. Among the materials used for this purpose, the alloys with shape memory effect, especially nitinol-alloy NiTi [1] are considered as the most promising materials [2-5].

Depending on the concentration of nickel in the alloy, Curie temperature of nitinol can cover the whole region of water temperatures in liquid state. Besides, the temperature hysteresis, e.g., of nitinol Ni(45%)Ti(55%) is equal to

$$\delta T = A_{fin} - M_{fin} = 30^0 C, \qquad (1)$$

where $M_{fin}$ and $A_{fin}$ are the final temperatures of direct austenite-martensite (A→M) and reverse martensite-austenite (M→A) transitions, respectively. All this, allows the nitinol engines to operate on heat energy of low-potential (LP) waters, among which are the geothermal waters, waste waters of industrial enterprises, waters heated by solar radiation, warm ocean streams, etc. There are also other martensite alloys with $\delta T$=4, 6, $10^0$C [6, 7] that can be used for conversion of heat energy of LP waters. Though, at present, there does not exist yet the efficient model of martensite heat engine providing the acceptable power and a rather high efficiency. The idea of creation of such engine remains still urgent dictated by the importance of searching for new nontraditional sources of energy and by the prospective production of ecologically pure electric energy at the expense of heat energy of LP waters, the resources of which is practically inexhaustible in the world.

In the present paper the mechanical and thermo-dynamic characteristics of MRHE are discussed, operating on the basis of structural phase transition M ↔ A with Curie temperature in the region of LP water temperatures. At some simplifying assumptions, the analytical expressions are obtained for the thermal efficiency and the power of MRHE of different type. The paper contains as well the recommendations on the optimal choice of engine parameters.

# 2. The design and the principle of MRHE operation

The martensite rotor heat engines [8, 9, 2-4] are superior to non-rotor martensite heat engines [10, 6, 7] in efficiency.

The design of MRHE is schematically shown in Fig. 1. The engine consists of two working pulleys 1 and 2 with $R_1$ and $R_2$ radii. The closed loop made of martensite alloy with shape memory effect – the working body (WB) of engine – is pulled over the working pulleys. WB can be a ribbon of $h$ thickness and $d$ width. MRHE with such WB we will call the ribbon engine and designate by the index RB ribbon: MRHE$_{Rb}$. WB can contain *n* wires of $r_0$ radius. Such MRHE we will call the wire engine and designate by the index *Wr (Wire)*: MRHE $_{Wr}$. WB can be made out of cylindrical spring of tightly winded (with turns touching) wire of $r_0$ radius. Such MRHE we will call the spring engine and designate by the index *Sp (spring):* MRHE$_{Sp}$. WB can contain *m* springs with the radius of wire $r_0$ and with the coefficient of elasticity $\mu = R_w/r_0$, which is the ratio of the radius of coil $R_w$ to the radius of wire the spring is coiled of. MRHE$_{Sp}$ and MRHE$_{Wr,Rb}$ strongly differ in rigidity of WB. On the shafts of working pulleys the auxiliary pulleys (gears) 3 and 4 are fixed, which, by means of rigid gearing 5 (or by gearing with intermediate pinions) make the working pulleys rotate in one and the same direction with angular velocities being in definite relation with each other - in so-called differential constraint.

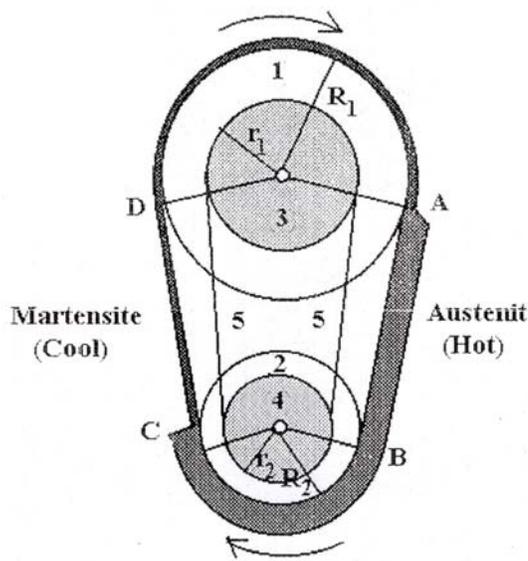

**Fig. 1. Schematic diagram of martensite rotor engine**

WB acts on the working pulleys with the circumferential force (CF) and make them rotate. One of the working pulleys, e.g. 1, is the driving pulley which rotates to the direction of CF action. In this case, pulley 2 is the driven pulley, as, owing to the differential constraint, it is forced to rotate opposite to CF acting on it.

Martensite alloy with the shape memory effect can be in two phase states: in more high-temperature austenite (A) and in more low-temperature martensite (M) states. Heating and cooling of WB are arranged in such a way that the temperature $A_{fin}$ of the end of phase transition M→A is achieved at point A (the point where WB runs down from pulley 1 in Fig.1) or earlier and is kept not lower than this level until WB runs on pulley 2 at point B. The temperature $M_{fin}$ of the end of phase transition A→M is achieved at point C (the point where WB runs down from pulley 2 in Fig.1) or earlier and is kept not higher than this level until WB runs on pulley 1 at point D. Correspondingly, during the operation of MRHE, WB consists of two branches – austenite (ABC section in Fig.1), which is heated continuously and is always in A, and martensite (CDA section in Fig.1), which is cooled continuously and is always in M. During the thermodynamic cycle, the martensite branch makes a full transition to austenite branch, and austenite – to martensite branch and is repeated cycle by cycle. Without loss of generality, it can be assumed that WB runs on pulley 1 in M and runs down from it in A, and vise versa, WB runs on pulley 2 in A and runs down from it in M, as it is shown in Fig. 1.

Calculation of MRHE parameters is made within the following approximations:

- skidding and elastic slip of WB are neglected; this problem does not generally arise, if CF is transferred to working pulleys not by friction but by gearing [3];

- at steady operation of MRHE, the distribution of deformations and tensions along WB does not depend on time, and the total length of WB does not change: the forced stretching of WB at phase transition A→M is fully compensated by contraction of WB at reverse phase transition M→A;

- phase transitions M→A and A→M take place in small neighborhood of A and C points (in the limit of A and C points) – ("narrow" phase transitions).

## 3. Cinematic characteristics of MRHE

Let us designate the gear ratio of differential constraint by *i*. Without the loss of generality, we can assume, as in Sec. 2, that the driving pulley 1 of WB runs on in M and runs down in A, and the driven pulley 2 of WB runs on in A and runs down in M. In this case, we have:

$$i = \omega_2 / \omega_1 = R_1 V_A / (R_2 V_M), \qquad (2)$$

where $\omega_{1,2}$ are the angular velocities of pulleys 1 and 2, and $V_{A,M}$ are the linear velocities of corresponding branches of WB. The choice of $i$ values is determined by the design of MRHE and by reasonability, e.g., in case of the known models of Johnson [8] and Patcher [9] $i \equiv 1$.

Let the temperature of WB be lower than the Curie temperature and WB be in martensite undeformed state and have the initial length $l_0$. (In case of cylindrical spring, this is the length of a spring wound with coils closing together: $l_0 = N \cdot 2r_0$, where $N$ is the number of coils, $r_0$ is the radius of wire the spring is coiled of). At the same temperature of WB, after it is pulled over the pulleys, its length increases and becomes equal to:

$$l = l_0(1 + \varepsilon_M) \qquad (3)$$

and should remain unchanged. $\varepsilon_M$ is the preliminary relative tensile deformation of WB in M, necessary for starting of MRHE, it is applied from outside and should satisfy the condition $\varepsilon_M \leq \varepsilon_{M,max}$, where $\varepsilon_{M,max}$ is the maximum permissible deformation of WB in M determined experimentally. At $\varepsilon_M > \varepsilon_{M,max}$ WB deformed in M is not fully recovered in A, i.e. it is damaged and MRHE does not work. If it is necessary to obtain the maximum power of MRHE, $\varepsilon_M = \varepsilon_M$ is chosen. But if a long-term operation of the engine is preferable, $\varepsilon_M < \varepsilon_M$ should be chosen.

The length $l$ of deformed WB is determined by geometrical dimensions of MRHE and depends on the radii of pulleys $R_1$ and $R_2$ and on the distance $d$ between the axes of rotation of pulleys. For the model shown in Fig. 1, we have

$$l = \pi(R_1 + R_2) + 2\sqrt{(R_1 - R_2)^2 + d^2} + 2(R_1 - R_2)\arcsin(R_1 - R_2)/d \qquad (4)$$

from which it follows that $l$ does not depend on the length of heated section $l_{ABC}$ of WB, i.e. on the length of austenite branch. At the given length $l$ and at the chosen value $\varepsilon_M$, the condition (3) determines the optimal initial length of WB.

$$\tilde{l}_0 = l/(1 + \varepsilon_M), \qquad (5)$$

which provides the maximum power at the given conditions. Incorrect choice of $l_0 \neq \tilde{l}_0$ leads either to the decrease of possible power of MRHE ($l_0 > \tilde{l}_0$), or to more rapid damage of WB ($l_0 < \tilde{l}_0$).

Before starting of MRHE, the austenite and martensite branches of WB are in M and have the lengths

$$l_{CDA} = \pi R_1 + 2\sqrt{(R_1 - R_2)^2 + d^2} + 2R_1 \arcsin(R_1 - R_2)/d$$

$$l_{ABC} = \pi R_2 + 2\sqrt{(R_1 - R_2)^2 + d^2} + 2R_2 \arcsin(R_1 - R_2)/d \qquad (6)$$

thus, the condition $l = l_{A0} + l_{M0}$ is fulfilled automatically.

After starting of MRHE, A branch is continuously poured with lowpotential (hot) water and in the steady regime of engine operation contracts to the initial length $l_0$ (shape memory effect!):

$$l_A = l_M(1-\varepsilon_A) = l_0, \quad (7)$$

where

$$l_M = l_0(1+\varepsilon_M) \quad (8)$$

is the length of M branch at the moment A branch starts to contract.

As the pulleys hinder the recovery of shape (compression!), the stresses are developed in WB, significantly exceeding the stresses caused by the initial deformation $\varepsilon_M$. As a result, the working pulleys are under the continuous action of CF, $\Delta F$, and the engine makes a useful work at the expense of heat energy of lowpotential water. Using (7) and (8), we have the important relation between the relative deformations of WB in A and in M:

$$\varepsilon_A = \varepsilon_M /(1+\varepsilon_M). \quad (9)$$

Let us use the condition of constancy of the flow of WB mass

$$\rho_M V_M = \rho_A V_A = \text{const}, \quad (10)$$

where $\rho_{M,A} = m_0/l_{M,A}$ are the linear densities of WB in M and in A, and $m_0$ is the mass of WB. From exprs. (7), (8) and (10) we find

$$V_A/V_M = (1+\varepsilon_M)^{-1}. \quad (11)$$

Substituting (11) in (2) we will obtain the necessary relation between the radii of working pulleys:

$$R_1 = R_2(1+\varepsilon_M)\cdot i. \quad (12)$$

## 4. Calculation of circumferential force

Circumferential force is equal to $\Delta F \sim F_A(\varepsilon) - F_M(\varepsilon)$, where $F_{AM}(\varepsilon)$ are the tensile forces of WB in A and M and $F_A > F_M$, while in case of MRHE$_{Wr, Rb}$ $F_A >> F_M$. Let us assume that the diagrams $F_A = F_A(\varepsilon)$ and $F_M = F_M(\varepsilon)$ are known from the experiment.

For the sake of simplicity, we assume that in the working interval of deformations, these diagrams can be changed by approximated linear dependences:

$$F_A = K_A \varepsilon_A, \quad F_M = F_{M0} + K_M \varepsilon_M, \quad (13)$$

where $K_A$, $K_M$, $F_{Mo}$ are the force constants that characterize WB material in A and M and are determined experimentally.

Heating of WB takes place in M, and cooling – in A, therefore, the driving pulley 1 rotates to the direction of circumferential force (CF) –$F_A$-$F_M$>0, and the driven pulley is forced to rotate opposite to this force thanks to the differential constraint. As a result, the torsional moment of MRHE appears to be equal to

$$M = (F_A - F_M)(R_1 - iR_2) = \Delta F \cdot R_1.$$

Using exp. (12) we find

$$\Delta F = M / R_1 = (F_A - F_M)\varepsilon_M (1+\varepsilon_M)^{-1}. \tag{14}$$

Let us consider the inevitable losses of energy due to the friction of elastic slip on pulleys, which are proportional to the relative loss of velocity $V_A/V_M$ which, according to (11), equals $V_A/V_M = (1+\varepsilon_M)^{-1}$. For this purpose, it is necessary to introduce an additional multiplier $(1+\varepsilon_M)^{-1}$ into $\Delta F$, and then CF will be determined by

$$\Delta F = (F_A - F_M)\varepsilon_M /(1+\varepsilon_M)^2. \tag{15}$$

Thus, the differential constraint providing the continuous motion of WB decreases CF significantly (multiplier $\varepsilon_M/(1+\varepsilon_M)$<<1 in (15)), as if "paying" the nature for the possibility of operation of MRHE. Using the approximation (13) and exp. (9), we obtain the final expression for CF

$$\Delta F = \left[(K_A - K_M - F_{M0})\varepsilon_M - F_{M0} - K_M\varepsilon_M^2\right]\varepsilon_M (1+\varepsilon_M)^{-3}. \tag{16}$$

Due to small deformations, for MRHE$_{Rb, Wr}$ the following conditions are satisfied well

$$F_A >> F_M \Rightarrow K_A >> K_M, F_{M0}; K_A\varepsilon_M >> F_{M0}, \tag{17}$$

therefore, CF for ribbon and wire MRHE is given by

$$\Delta F_{Rb,Wr} = K_{ARb,Wr}\varepsilon_M^2 (1+\varepsilon_M)^{-3}. \tag{18}$$

Let us estimate $\Delta F_{Sp}$. As, for the operation of MRHE$_{Sp}$ it is necessary to meet the condition $\Delta_F$>0, we obtain the limitation on $\varepsilon_M$:

$$\bar{\varepsilon} - \sqrt{\bar{\varepsilon}^2 - \Delta} < \varepsilon_M < \bar{\varepsilon} + \sqrt{\bar{\varepsilon}^2 - \Delta},$$

where

$$\bar{\varepsilon} = (K_A - K_M - F_{M0})\cdot(2K_M)^{-1}, \quad \Delta = F_{M0}/K_M. \tag{19}$$

At the given parameters (17) we find $\widetilde{\varepsilon}_M \cong 7$, 0.3<$\varepsilon_M$<13.7

The force constants of spring made of nitinol Ni(45%)Ti(55%) known from the experiment, are equal to

$$F_{M0} \approx 2H, \quad K_M \approx 0.44H, \quad K_A \approx 8.6H$$

therefore, the limitation (19) leads to the estimation of $0.3 < \varepsilon_M < 13.7$.

From the condition $\dfrac{\partial \Delta F_{Sp}}{\partial \varepsilon_M} = 0$ it is easy to show, that $\Delta F_{Sp,\max} \cong 0.63 N$ at $\varepsilon_M = 2$, besides, the width of peak of $\Delta F_{Sp}(\varepsilon_M)$ function is rather large (of the order of 1-2) caused by a large permissible deformation of spring. This gives the possibility to select empirically the necessary value of $\varepsilon_M$ and acceptable ratio between the radii of pulleys $R_1$ and $R_2$ when designing MRHE$_{Sp}$.

For calculation of $\Delta F_{wr}$, let us use exp.(18) and the experimental value $K_{AMr} \cong 3.4 \cdot 10^4 \, N$. Giving the value of initial deformation $\varepsilon_M = 0.015 = 1.5\%$ (the reason of such choice is given below), by exp. (22) we estimate $\Delta F_{Wr} \cong 7.3 N$, exceeding CF of MRHE$_{Sp}$ by more than an order of magnitude.

## 5. Calculation of thermal efficiency and power of MRHE

MRHE is "feeding" by the heat energy of LP water by means of heat transfer at the boundary of WB – LP water. Let us make this process similar to heat transfer of metal surface at the boundary with the moving water. Then, the intensity of the flow of heat transfer will be equal to [11]

$$W = \left(\sigma_0 + \sigma_1 \sqrt{V_{rel}}\right) S_W \Delta T, \qquad (20)$$

where $V_{rel} = V_M$ is the velocity of heated section of WB (ribbon) with the length of $l_w$ relative to water, $\Delta T = T_{warm} - T_{cool}$ is the temperature gradient (thermal head) – the difference between the temperatures of LP water, $T_{warm}$ (hot) and $T_{cool}$ (cold) water, and $\sigma_0$, $\sigma_1$ are the coefficients of heat transfer equal to

$$\sigma_0 = 350 \, Bm/(m^2 K), \; \sigma_1 = 2100 \, Bm/(m^2 K) \cdot (c/m)^{1/2}. \qquad (21)$$

The contact surface $S_w$ of WB with LP water for MRHE of different kind is determined by

$$S_W(Rb) = 2dl_W, \; S_W(Wr) = n 2\pi r_0 l_W, \; S_W(Sp) = 4\pi r_0 (\mu - 1) l_W, \qquad (22)$$

Apparently, W is the sum of two summands:

$$W = W_Q + W_A, \qquad (23)$$

where $W_Q$ is a part of W consumed continuously for heating of WB at inverse martensite transition M→A, and $W_A$ is a part of W at the expense of which MRHE develops the power N and at this transition it does the work A:

$$A = \Delta F \cdot l_W = N \tau_Q, \qquad (24)$$

where $\tau_Q$ is the time of heating equal to

$$\tau_Q = l_W / V_M = Q / W_Q = A / N. \tag{25}$$

Here, Q is the quantity of heat transferred to WB for $\tau_Q$ time at steady operation of MRHE, necessary for M→A transition

$$Q = (\overline{C\rho}\,\delta T + \Delta q)V_W. \tag{26}$$

The value $\overline{C\rho}$ is the average value of the product of specific heat capacity of nitinol $C$, the material of WB, and of its density – $\rho$ $\Delta_q$ is the latent heat of transition per volume unit, $V_w$ is the volume of the heated part of WB determined by

$$V_W(Rb) = dhl_W, \quad V_W(Wr) = n\pi r_0^2 l_W, \quad V_W(Sp) = m\pi^2 \mu r_0^2 l_W \tag{27}$$

From the experiment it is known that in the case of nitinol

$$\delta T = 30K, \; \overline{C\rho} = 5.2 \cdot 10^6 \, J/(m^3 K), \tag{28}$$

i.e. $\overline{C\rho}\,\delta T \gg \Delta q$, therefore

$$Q = \overline{C\rho}\,\delta T \cdot V_W. \tag{29}$$

By definition, the thermal efficiency is the relation between the work and the quantity of heat received by WB at heating, i.e.

$$\eta = A/Q = \Delta F \cdot l_W / Q = \Delta F \cdot l_W / (\overline{C\rho}\,\delta T \cdot V_W). \tag{30}$$

The thermal efficiency of any heat engine is limited from "above" $\eta \leq \eta_K$, where $\eta_K$ is the efficiency of an ideal heat engine. Under the conditions of LP water, we, as an example, take $T_{warm}=313K$, $T_{cool} = 283K$. Then, $\eta_K=(T_{warm}-T_{cool})/T_{warm}\cdot 100\% \cong 9.6\%$. In accordance with (30) and (18), this means that MRHE$_{Wr, Rb}$ can operate only at relative deformations $\varepsilon_M \leq \varepsilon_{MK}=1.65\%$. Just for this reason, at the estimation of $\Delta F_{Wr}$ in the previous section we took the value $\varepsilon_M=1.5\%$, which is much less than $\varepsilon_{M, max}$. (In the case of MRHE$_{Sp}$ "thermodynamic" limitation on $\eta_{Sp}$ is insignificant due to large permitted deformations).

Let us assume $Q=W_Q\tau_Q$, $l_W=V_M\tau_Q$ according to (25) and substitute these values into (30), then we obtain the relation $W_Q=N/\eta$.

On the other hand, the quantity of heat $W_A\tau_Q$ is consumed for execution of work A, therefore, the thermal efficiency η can be presented in the form equivalent to (30)

$$\eta = \frac{A}{W_A \tau_q} = \frac{N}{W_A}, \tag{31}$$

from which it follows that

$$W_Q = W_A = N/\eta. \tag{32}$$

By substitution of (32) into (23) we obtain one of the main relations of phenomenological theory of MRHE

$$N = \eta W / 2 \tag{33}$$

Thus, the power of MRHE is determined trivially by calculation of W using exp. (20), and of η using exprs. (30), (29), (16) and (19). Thanks to the existence of feedback N→W in the cycle N→$V_M$→N→W, MRHE can reach the significant powers. The relative infinitesimal η obtained from estimations is not a determining value, as MRHE operates at restored energy of LP water.

Substitution of (33) into (32) gives an interesting result

$$W_Q = W_A = W/2. \tag{34}$$

At the first glance, the obtained relation seems unexpected. But, in fact, there is nothing surprising. The heating of WB at the expense of heat and the execution of work at the expense of heat $W_A\tau_Q$ take place simultaneously during the time $\tau_Q$, therefore, in steady regime the following condition should be fulfilled (34).

From (33) it follows that the coefficient of heat transfer is equal to N/W=η/2. Using (25), (32) and (29) we easily find that at N power the engine requires the volumetric water discharge, which, according to (24) and (25) is:

$$\tau_Q = N V_W / (\Delta F l_W). \tag{35}$$

As $V_W \sim l_W$, from (30) and (35) it follows that the efficiency and the volumetric water discharge do not depend on the length of the heated part of WB of engine.

Let us transform the expression for power (33) to the view convenient for analysis. Substituting (20), (24) into (33), we obtain the following expression for the power of MRHE of any kind:

$$N = N_0 (1 + \sigma_1 \sqrt{N/\Delta F} / \sigma_0), \tag{36}$$

where the following notation is introduced:

$$N_0 = \frac{1}{2} \eta \sigma_0 S_W \Delta T. \tag{37}$$

From (36) we can easily obtain the quadratic equation for determination of N, the root of which satisfying the apparent condition N>$N_0$, has the following form:

$$N = N_0 \left[1 + P + \sqrt{P(P+2)}\right], \tag{38}$$

where

$$P = \sigma_1^2 N_0 / (2\sigma_0^2 \Delta F) \tag{39}$$

is the dimensionless thermodynamic parameter determining the power of MRHE.

N is the monotonously increasing function of *P* parameter, therefore, for obtaining the maximum power it is necessary to take the values of P as high as possible.

The power of MRHE (38) is calculated without consideration of hydraulic resistance that imposes the limitation on N from "above".

## 6. Conclusion

Accounting of the hydraulic resistance, heat conductivity of working body material, conditions of exit MRHE on maximal power and comparison with experiment will be published in next article.

**Acknowledgements:** This work is supported by the Shota Rustaveli National Science Foundation, Grant GNSF 712/07. The authors are grateful to M. Nikoladze for help with processing articles